\documentclass[a4paper]{article}
\usepackage{amsfonts}
\usepackage{amsmath}

\input{tcilatex}

\begin{document}

\title{Application of Hellman-Feynman and Hypervirial theorems to the
eigenvalue problem: Coulomb plus linear term and quartic anharmonic
oscillator potentials}
\author{S.Rekab and N.Zenine \\
Centre de Recherche Nucl\'{e}aire d'Alger, \\
02 Bd Frantz Fanon, Alger, Algeria\\
e-mail : srekab@comena-dz.org, nzenine@comena-dz.org}
\maketitle

\begin{abstract}
We use the Hellman-Feynman (HF) and Hypervirial (HV) theorems, to calculate
the perturbative coefficients of the eigenenergies formal series, in the
case of the Coulomb potential with a radial linear term and the radial
quartic anharmonic oscillator potential. This calculation method, contrary
to the usual Rayleigh-Schr\"{o}dinger Perturbation Theory (RSPT), does not
require the calculation of eigenfunctions coefficients. This method is a
fast and efficient tool for the calculation of large order eigenenergies
coefficients.
\end{abstract}

\section{\protect\bigskip Introduction}

Many physical phenomena encountered in different fields of physics, require
the modelization of the interaction through potentials that are believed to
account for the physical mechanisms occurring in the system. The
construction of these potentials is constrained by physical consideration in
their analytical forms and introduces parameters. For instance, the problem
of the electron in the field of a nucleus is tackled with a Coulomb
potential, since it is known that this is the interaction which is in work.
In nuclear physics, the nuclear energy levels are obtained by modelizing the
interaction, the nucleon feels through (e.g.) a Woods-Saxon potential. In
particle physics, the mesons are described as quark antiquark bound states
through a phenomenological model where the potential is given at once.

While bound states problem can be solved analytically \cite{flugge} for few
potentials, among which Coulomb and harmonic oscillator potentials, for most
of potentials of physical interest, one has to resort to direct numerical
techniques or to approximation methods. Note that, beyond the physical
problem itself, these solving techniques have become a field of study in
their own.

Perturbation theory is one of the most widely used approximation method to
solve eigenvalue problems in theoretical physics \cite{wilcox, killingbeck,
orzagbender}. Mathematical aspects of perturbation theory are treated in
monographs \cite{maslov, kato, reedsimon}, while applications in various
fields are described in books \cite{kumar, clark, adams}.

Rayleigh-Schr\"{o}dinger Perturbation Theory (RSPT) is designed for systems
whose Hamiltonian can be partitioned into a \emph{completely solvable }part
with known eigenvalues and eigenfunctions, and a perturbation part \cite%
{wilcox, killingbeck, orzagbender, MQ}. The formalism of RSPT expresses the
eigenenergies and the eigenfunctions as \emph{formal series} in the choosen
perturbation parameter. Accurate results for physical observables have been
obtained from this perturbative method. However, one of the shortcomings of
the RSPT approach is that it requires, in general, considerable
computational time and effort. Indeed, this method uses the eigenfunctions
solutions which are time consuming and intermediary results in an eigenvalue
problem \cite{orzagbender, MQ}.

For various interactions, a way to bypass the computation of the
eigenfunctions is to use Hellman-Feynman (HF) and the Hypervirial (HV)
theorems \cite{HFT, HVT} to generate eigenenergies perturbative coefficients 
\cite{kill78, grant79, fernandez87, vrscay93, kwato01}.

The purpose of the paper is to use these theorems (denoted by HFHV method)
to calculate the eigenvalues perturbative coefficients for two potentials of
physical interest namely: the Coulomb potential plus radial linear term and
the radial quartic anharmonic oscillator potential. The eigenenergies
coefficients are computed recursively without the need to calculate the
eigenfunctions. Moreover, we show that the application of the HFHV method
requires considerable less computational time comparatively to the usual
RSPT method and allows a fast and efficient calculation of very large order
eigenvalues coefficients.

\section{Method of calculation}

\subsection{Hellman-Feynman and Hypervirial theorems}

Let us recall Hellman-Feynman (HF) and Hypervirial (HV) theorems. The HF
theorem states that if we consider a system with a Hamiltonian $H=H(\lambda
) $, depending on some parameters $\lambda$, with normalized eigenvector $%
\Psi (\lambda )$ and eigenvalue $E(\lambda )$, then: 
\begin{equation}
\ \ \frac{\partial E}{\partial \lambda }=\langle \Psi (\lambda )\left\vert 
\frac{\partial H}{\partial \lambda }\right\vert \Psi (\lambda )\rangle
=\langle \frac{\partial H}{\partial \lambda }\rangle  \label{HF}
\end{equation}

Let $H$ be a time-independant Hamiltonian with a normalized bound state
eigenfunction $\Psi $, and $\Theta $ a time-independant linear operator. If
the expectation value of $\Theta $ over the stationnary eigenstate $\Psi $ 
\emph{is not infinite}, then the HV theorem states that, for the operator $%
\Theta $: 
\begin{equation}
\langle \Psi \mid \lbrack \Theta ,H]\mid \Psi \rangle =\langle \text{ }%
[\Theta ,H]\text{ }\rangle =0  \label{HV}
\end{equation}%
where $[\Theta ,H]=\Theta H-H$ $\Theta $ is the commutator of $H$ and $%
\Theta $. Physically, this, is just the statement that, for a stationary
state, the expectation value of $\Theta $ is time-independant.

\subsection{General formalism}

The two-body stationnary Schr\"{o}dinger equation with central potential $%
V(r)$, can be written (in $\hbar =c=1$ units ) as: 
\begin{equation}
\left[ -\frac{1}{2\mu }\frac{d^{2}}{dr^{2}}+V(r)+\frac{l(l+1)}{2\mu r^{2}}%
-E_{n,l}\right] \text{ \ }y_{n,l}(r)=0  \label{sch0}
\end{equation}%
where $\mu $ is the reduced mass of the system, $E_{n,l}$ and $y_{n,l}(r)$
are respectively the eigenvalue and the reduced radial wave function,
describing a bound state of radial and orbital angular momentum quantum
numbers $n$ and $l$ respectively. The wave function obeys the following
boundary conditions: 
\begin{equation}
y_{n,l}(0)=y_{n,l}(\infty )=0\text{ }  \label{boundcond}
\end{equation}%
Setting $x=\lambda r$, the scaling property of the Schr\"{o}dinger equation
allows to write (\ref{sch0}) as: 
\begin{equation}
H\ y_{n,l}(x)=\epsilon _{n,l}\ y_{n,l}(x)  \label{sch1}
\end{equation}%
with: 
\begin{equation}
H=-\frac{d^{2}}{dx^{2}}+U(x)  \label{ham0}
\end{equation}%
and: 
\begin{equation}
U(x)\,=\,V_{0}(x)+\frac{l(l+1)}{x^{2}}+g\,V_{1}(x)  \label{Uxtot}
\end{equation}%
The perturbation is cast in the term $g\,V_{1}(x)$ where $g$ is the
controlling parameter, i.e., the expansion parameter.

In the usual RSPT method, the eigenvalues and eigenfunctions are written as
series expansions in the perturbative parameter $g$: 
\begin{equation}
\epsilon _{n,l}=\sum_{k=0}^{\infty }\epsilon _{n,l}(k)\text{ \ }g^{k}
\label{EVformal}
\end{equation}%
\begin{equation}
y_{n,l}(x)=\sum_{k=0}^{\infty }y_{n,l}(k,x)\text{ \ }g^{k}  \label{EFformal}
\end{equation}%
where $\epsilon _{n,l}(k)$ and $y_{n,l}(k,x)$ are, respectively, the
eigenenergies and eigenfunctions perturbative coefficients. $k=0$ in the
expansion corresponds to the unperturbed system, i.e., $\epsilon _{n,l}(0)$
and $y_{n,l}(0,x)$ are the eigenvalues and eigenfunctions for the potential (%
\ref{Uxtot}) with $g=0$.

With the RSPT method, the calculation of $\epsilon _{n,l}(k)$ coefficients
requires then the knowledge of $y_{n,l}(i,x)$ where $i=0,\cdots k-1$ \cite%
{wilcox, killingbeck, orzagbender, MQ}. This is the shortcomings of the
method.

It happens that, for various interactions, and thanks to HF and HV theorems,
the $\epsilon _{n,l}(k)$ coefficients can be given by a recursion formula
(see below) and thus there is no need to calculate the intermediary
eigenfunctions \cite{kill78, grant79, fernandez87, vrscay93, kwato01}.

\subsection{The potentials}

We consider two central potentials of physical interest, the Coulomb
potential with a radial linear term and the quartic anharmonic oscillator
potential.

The Coulomb potential with a radial linear term: 
\begin{equation}
V(r)=-\frac{\alpha }{r}+\beta r\qquad \alpha ,\beta \geq 0  \label{pot1}
\end{equation}%
corresponds to a spherical Stark effect in hydrogenic atoms and is, also, a
successful phenomenological model for the description of heavy quark
antiquark bound states in particle physics \cite{cornell, lucha}. In this
case, taking $\lambda =2\mu \alpha $, the eigenvalues are given by: 
\begin{equation}
\epsilon _{n,l}=\frac{1}{\alpha ^{2}}\frac{E_{n,l}}{2\mu }  \label{eigval1}
\end{equation}%
and the scaled potential (\ref{Uxtot}) reads: 
\begin{equation}
U(x)=-\frac{1}{x}+\frac{l(l+1)}{x^{2}}+g\cdot x,\qquad g=\frac{\beta }{(2\mu
)^{2}\alpha ^{3}}  \label{potU1}
\end{equation}%
where $g$ is the dimensionessless parameter used in the expansion. The term $%
g\,x$ is thus treated as a perturbation, while the the unperturbed
Hamiltonian is the usual radial Coulomb Hamiltonian whose eigenfunctions and
eigenvalues are known \cite{flugge, MQ}.

The second potential is the quartic anharmonic oscillator potential: 
\begin{equation}
V(r)=ar^{2}+br^{4}\text{ \ \ \ \ \ }a,b\geq 0  \label{pot2}
\end{equation}%
which has received a great deal of interest because of its importance in the
understanding of molecular vibrations \cite{molecvib}, in some areas of
solid state physics \cite{solide1,solide2} and in Quantum Field Theory \cite%
{benderwu69,chang75}. Furthermore, the quartic anharmonic oscillator serves
as a testing tool to check various approximative methods in Quantum
Mechanics \cite{mst, gta}. Setting $\lambda =(2\mu $ $a)^{1/4}$, the
eigenvalues are given by: 
\begin{equation}
\epsilon _{n,l}=\sqrt{\frac{2\mu }{a}}E_{n,l}  \label{eigval2}
\end{equation}%
and the scaled form of this potential (\ref{Uxtot}) reads: 
\begin{equation}
U(x)=x^{2}+\frac{l(l+1)}{x^{2}}+g\cdot x^{4},\qquad g=\frac{b}{\sqrt{2\mu }%
a^{3/2}}  \label{potU2}
\end{equation}%
where $g$ is the perturbative dimensionessless parameter. Here again, the
term $g\,x^{4}$ is treated as a perturbation, while the unperturbed
Hamiltonian is the radial harmonic oscillator Hamiltonian whose
eigenfunctions and eigenvalues are known \cite{flugge, MQ}.

\subsection{Recurrence equations on the $\protect\epsilon _{n,l}(k)$\
coefficients}

Let us consider the time-independant linear operator $\Theta $ as $\Theta
=x^{j+1}\frac{d}{dx}$, where $j$ is an integer such that $j\geq 0$ \cite%
{kill78, grant79, kwato01, killingbeck80, killingbeck2001}. Taking the
Hamiltonian in the form (\ref{ham0}), one has:%
\begin{equation}
\lbrack x^{j+1}\frac{d}{dx},H]=2(j+1)\text{ }x^{j}\text{ }\frac{d^{2}}{dx^{2}%
}+j(j+1)x^{j-1}\text{ }\frac{d}{dx}+x^{j+1}\text{ }\frac{dU(x)}{dx}\text{ }
\label{commutator}
\end{equation}%
From the definition (\ref{sch1}), one can easily show that\footnote{%
Throughout this paper, the notation $<\xi >$ stands, the quantum number
indices $n$ and $l$ being understood, for the expectation value $<\xi
>=\langle \Psi \mid \xi \mid \Psi \rangle $, where $\Psi $ represents the
normalized eigenstate of the hamiltonian $H$ under consideration.}:%
\begin{equation}
\langle x^{j}\text{ }\frac{d^{2}}{dx^{2}}\rangle =\langle \text{ }x^{j}\text{
}U(x)\text{ }\rangle -\epsilon _{n,l}\text{ }\langle x^{j}\text{ }\rangle 
\label{valmoy1}
\end{equation}%
Integration by parts, using the boundary conditions (\ref{boundcond})
together with the asymptotic behavior $y_{n,l}(x)\sim x^{l+1}$ for small $x$%
, leads to:%
\begin{equation}
\langle x^{j-1}\text{ }\frac{d}{dx}\rangle =-\frac{1}{2}(j-1)\text{ }\langle 
\text{ }x^{j-2}\text{ }\rangle   \label{valmoy2}
\end{equation}%
Inserting relations (\ref{commutator}-\ref{valmoy2}) into equation (\ref{HV}%
), the HV theorem provides relationships between $\epsilon _{n,l}$ and the
various expectation values $\langle x^{j}$ $\rangle $ through the following
equations \cite{kill78, grant79, kwato01, killingbeck80, killingbeck2001}:%
\begin{equation}
2(j+1)\text{ }\epsilon _{n,l}\text{ }\langle x^{j}\text{ }\rangle =-\frac{1}{%
2}j(j^{2}-1)\langle \text{ }x^{j-2}\text{ }\rangle +2(j+1)\text{ }\langle 
\text{ }x^{j}\text{ }U(x)\text{ }\rangle +\langle \text{ }x^{j+1}\text{ }%
\frac{dU(x)}{dx}\text{ }\rangle   \label{HVeq0}
\end{equation}%
The HF theorem (\ref{HF}) allows to derive a further relation:%
\begin{equation}
\frac{\partial \epsilon _{n,l}}{\partial g}=\langle \frac{\partial U(x)}{%
\partial g}\rangle =\langle \text{ }V_{1}(x)\text{ }\rangle =\left\{ 
\begin{array}{c}
\langle x\rangle \hbox{\,\,for the potential\,\,}(\ref{potU1}) \\ 
\langle x^{4}\rangle \hbox{\,\, for the potential\,\,}(\ref{potU2})%
\end{array}%
\right\}   \label{HFeq0}
\end{equation}%
The essence of the application of the HF and HV theorems (HFHV perturbative
method) is to assume, in the spirit of perturbative methods, that the energy
and the expectation values of position coordinates can be expanded in power
series of the perturbation parameter $g$ as (see eq. (\ref{EVformal}) for
the energy):%
\begin{equation}
\langle \text{ }x^{j}\text{ }\rangle =\sum_{k=0}^{\infty }x_{j}^{(k)}\text{
\ }g^{k}  \label{xjexp0}
\end{equation}%
By equating like powers of $g$ on both sides of equations (\ref{HVeq0}-\ref%
{HFeq0}) after substitution of expansions (\ref{EVformal}) and (\ref{xjexp0}%
), we find a set of coupled relations involving $\epsilon _{n,l}(k)$ and $%
x_{j}^{(k)}$.

To be more precise, let us consider the Cornell potential defined by (\ref%
{pot1}) or (\ref{potU1}). In this case, equation (\ref{HVeq0}) can be
written as:%
\begin{equation}
\text{ }\epsilon _{n,l}\text{ }\langle x^{j}\text{ }\rangle =\alpha _{j}%
\text{ }\langle \text{ }x^{j-2}\text{ }\rangle +\beta _{j}\text{ }\langle 
\text{ }x^{j-1}\text{ }\rangle +g\text{ }\gamma _{j}\text{ }\langle \text{ }%
x^{j+1}\text{ }\rangle  \label{HVpot1}
\end{equation}%
with:%
\begin{equation}
\alpha _{j}=j\text{ }\frac{\left( (2l+1)^{2}-j^{2}\right) }{4j+4};\text{ \ \ 
}\beta _{j}=-\text{ }\frac{2j+1}{2j+2};\text{ \ \ }\gamma _{j}=\text{ }\frac{%
2j+3}{2j+2}  \label{albegam1}
\end{equation}%
Inserting the expansions (\ref{EVformal}) and (\ref{xjexp0}), in equations (%
\ref{HVpot1}-\ref{HFeq0}), the identification of the coefficients of powers
of $g$ leads to ($j$ integer $\geq 0$):%
\begin{equation}
\sum_{p=0}^{k}\epsilon _{n,l}(k-p)\text{ \ }x_{j}^{(p)}=\alpha _{j}\text{ }%
x_{j-2}^{(k)}+\beta _{j}\text{ }x_{j-1}^{(k)}+\theta (k-1)\text{ }\gamma _{j}%
\text{ }x_{j+1}^{(k-1)}\text{ \ \ \ \ }k\geq 0  \label{rec1a}
\end{equation}%
\begin{equation}
\epsilon _{n,l}(k)=\frac{x_{1}^{(k-1)}}{k}\text{ \ \ \ \ \ \ \ \ \ \ \ \ \ \
\ }k\geq 1  \label{rec1b}
\end{equation}%
where $\theta (t)$ is the Heaviside Step function. It is clear from the
above equations that the starting point is the normalization condition:%
\begin{equation}
x_{0}^{(0)}=1\text{ \ \ and \ }x_{0}^{(k)}=0\text{ \ \ for\ \ }k\geq 1
\label{normx0}
\end{equation}%
and the Coulomb eigenvalues:%
\begin{equation}
\epsilon _{n,l}(0)\text{ \ }=-\frac{1}{4(n+l+1)^{2}}  \label{cbeigen0}
\end{equation}%
The set of equations (\ref{rec1a}-\ref{rec1b}) together with (\ref{normx0}-%
\ref{cbeigen0}) allows one (see below) to calculate all the $\epsilon
_{n,l}(k)$ coefficients.

In a similar way for the quartic anharmonic oscillator (\ref{pot2}) or (\ref%
{potU2}), taking $j$ even ($j=2i$), equation (\ref{HVeq0}) reads:%
\begin{equation}
\text{ }\epsilon _{n,l}\text{ }\langle x^{2i}\text{ }\rangle =\alpha _{i}%
\text{ }\langle \text{ }x^{2i-2}\text{ }\rangle +\beta _{i}\text{ }\langle 
\text{ }x^{2i+2}\text{ }\rangle +g\text{ }\gamma _{i}\text{ }\langle \text{ }%
x^{2i+4}\text{ }\rangle  \label{HVpot2}
\end{equation}%
with:%
\begin{equation}
\alpha _{i}=i\text{ }\frac{\left( (2l+1)^{2}-4i^{2}\right) }{4i+2};\text{ \
\ }\beta _{i}=\text{ }\frac{2i+2}{2i+1};\text{ \ \ }\gamma _{i}=\text{ }%
\frac{2i+3}{2i+1}  \label{albegam2}
\end{equation}%
By analogy with equations (\ref{rec1a}-\ref{rec1b}) and noting $x_{2i}^{(p)}=%
\widehat{x}_{i}^{(p)}$, one obtains in the quartic anharmonic potential case:%
\begin{equation}
\sum_{p=0}^{k}\epsilon _{n,l}(k-p)\text{ \ }\widehat{x}_{i}^{(p)}=\alpha _{i}%
\text{ }\widehat{x}_{i-1}^{(k)}+\beta _{i}\text{ }\widehat{x}%
_{i+1}^{(k)}+\theta (k-1)\text{ }\gamma _{i}\text{ }\widehat{x}_{i+2}^{(k-1)}%
\text{ \ \ \ \ }k\geq 0  \label{rec2a}
\end{equation}%
\begin{equation}
\epsilon _{n,l}(k)=\frac{\widehat{x}_{2}^{(k-1)}}{k}\text{ \ \ \ \ \ \ \ \ \
\ \ \ \ \ \ }k\geq 1  \label{rec2b}
\end{equation}%
together with the normalization condition:%
\begin{equation}
\widehat{x}_{0}^{(0)}=1\text{ \ \ and \ }\widehat{x}_{0}^{(k)}=0\text{ \ \
for\ \ }k\geq 1  \label{normx02}
\end{equation}%
and the harmonic oscillator eigenvalues:%
\begin{equation}
\epsilon _{n,l}(0)\text{ \ }=3+4n+2l  \label{harmeigen0}
\end{equation}

\subsection{Calculation of the $\protect\epsilon _{n,l}(k)$\ coefficients}

To calculate explicitely the $\epsilon _{n,l}(k)$\ coefficients from the
equations (\ref{rec1a}-\ref{rec1b}) or (\ref{rec2a}-\ref{rec2b}), one
proceeds in a hierarchical manner by giving to $k$ various integer values
starting with $k=0$. Let us describe this method in details for the case of
the Cornell potential. Consider equation (\ref{rec1a}) for $k=0$, one
obtains:%
\begin{equation}
\epsilon _{n,l}(0)\text{ \ }x_{j}^{(0)}=\alpha _{j}\text{ }%
x_{j-2}^{(0)}+\beta _{j}\text{ }x_{j-1}^{(0)}  \label{rec1ak=0}
\end{equation}%
which is a recurrence of depth three on the \ $x_{j}^{(0)}$\ coefficients.
One needs two initial conditions to calculate all the \ $x_{j}^{(0)}$\
coefficients. One condition is given by the normalization (\ref{normx0}),
namely $x_{0}^{(0)}=1$, the secund one is given by taking \ $j=0$ in eq. (%
\ref{rec1ak=0}) and one has:%
\begin{equation}
\text{\ }x_{-1}^{(0)}=-2\text{ }\epsilon _{n,l}(0)\text{ }=\frac{1}{%
2(n+l+1)^{2}}  \label{cond1k=0}
\end{equation}%
The recurrence equation (\ref{rec1ak=0}) together with the initial
conditions (\ref{normx0}) and (\ref{cond1k=0}) allow to calculate all $%
x_{j}^{(0)}$\ coefficients. The next step is to consider equation (\ref%
{rec1a}) for $k$ $\geq 1$, which can be written, using (\ref{rec1b}), in the
following form:%
\begin{equation}
\epsilon _{n,l}(0)\text{ \ }x_{j}^{(k)}=\alpha _{j}\text{ }%
x_{j-2}^{(k)}+\beta _{j}\text{ }x_{j-1}^{(k)}+\delta (j,k)\text{ \ \ \ \ \ }%
k\geq 1  \label{rec1aksup1}
\end{equation}%
where $\delta (j,k)$ depends only on \ $x_{j}^{(p)}$ terms with $p<k$ and
reads:%
\begin{equation}
\delta (j,k)=-\sum_{p=0}^{k-1}\text{ }\frac{x_{1}^{(k-1-p)}\text{\ }%
x_{j}^{(p)}}{k-p}+\text{ }\gamma _{j}\text{ }x_{j+1}^{(k-1)}\text{ }
\label{inh1}
\end{equation}%
The obtained equation (\ref{rec1aksup1}) is an $\emph{inhomogenous}$
recurrence of depth three on the \ $x_{j}^{(k)}$\ coefficients ($k$ kept
fixed), $\delta (j,k)$ being the $\emph{inhomogenous}$ term. In the same
way, one needs two initial conditions to calculate the \ $x_{j}^{(k)}$\
coefficients. One condition is given by the normalization (\ref{normx0}),
namely $x_{0}^{(k)}=0$ ($k$ $\geq 1$), the secund one is given by taking \ $%
j=0$ in eq. (\ref{rec1aksup1}) and one has, using (\ref{normx0}):%
\begin{equation}
x_{-1}^{(k)}=\left( 3-\frac{2}{k}\right) \text{ }x_{1}^{(k-1)}\text{ \ \ \ \ 
}k\geq 1  \label{cond1ksup1}
\end{equation}%
One proceeds step by step by giving to $k$ various integer values in
equations (\ref{rec1aksup1}-\ref{cond1ksup1}), starting with $k=1$.

Finally , we can, thus, calculate the $r_{0}^{\text{th}}$ perturbative
coefficient $\epsilon _{n,l}(r_{0})$ through the relation (\ref{rec1b}) from
the knowledge of the \ $x_{j}^{(k)}$ with\footnote{%
For $j=-1,0$, the $x_{j}^{(k)}$ are given by the initial conditions (\ref%
{normx0}, \ref{cond1k=0} and \ref{cond1ksup1}) and the recurences displayed
above allow to calculate the $x_{j}^{(k)}$ for $j\geq 1$.} $0\leq k\leq
r_{0}-1$ and $-1\leq j\leq r_{0}-k$.

For the quartic anharmonic oscillator (\ref{potU2}), similarly, one obtains (%
$k=0$):%
\begin{equation}
\beta _{i}\text{ }\widehat{x}_{i+1}^{(0)}=\epsilon _{n,l}(0)\text{ \ }%
\widehat{x}_{i}^{(0)}-\alpha _{i}\text{ }\widehat{x}_{i-1}^{(0)}
\label{rec2ak=0}
\end{equation}%
\begin{equation}
\widehat{x}_{0}^{(0)}=1\text{ \ and \ }\widehat{x}_{1}^{(0)}=\epsilon
_{n,l}(0)/2=3/2+2n+l\text{ }  \label{cond2k=0}
\end{equation}%
and ($k$ $\geq 1$):%
\begin{equation}
\beta _{i}\text{ }\widehat{x}_{i+1}^{(k)}=\epsilon _{n,l}(0)\text{ \ }%
\widehat{x}_{i}^{(k)}-\alpha _{i}\text{ }\widehat{x}_{i-1}^{(k)}+\delta (i,k)%
\text{ }  \label{rec2aksup1}
\end{equation}%
\begin{equation}
\delta (i,k)=-\gamma _{i}\text{ }\widehat{x}_{i+2}^{(k-1)}+\sum_{p=0}^{k-1}%
\frac{\widehat{x}_{2}^{(k-1-p)}\text{ \ }\widehat{x}_{i}^{(p)}}{k-p}
\label{inh2}
\end{equation}%
\begin{equation}
\widehat{x}_{0}^{(k)}=0\text{ \ and \ }\widehat{x}_{1}^{(k)}=\frac{(-3+1/k)}{%
2}\widehat{x}_{2}^{(k-1)}  \label{cond2ksup1}
\end{equation}%
Consequently, the $r_{0}^{\text{th}}$ perturbative coefficient $\epsilon
_{n,l}(r_{0})$ is calculated through the relation (\ref{rec2b}) from the
knowledge of the \ $\widehat{x}_{j}^{(k)}$ with\footnote{%
For $j=0,1$, the $\widehat{x}_{j}^{(k)}$ are given by the initial conditions
(\ref{cond2k=0} and \ref{cond2ksup1}) and the recurences displayed above
allow to calculate the $\widehat{x}_{j}^{(k)}$ for $j\geq 2$.} $0\leq k\leq
r_{0}-1$ and $0\leq j\leq r_{0}-k+1$.

\section{Results and discussion}

We have implemented the HFHV perturbative method described above for the
calculations of $\epsilon _{n,l}(k)$ coefficients, \emph{at any order }$k$,
into a Mathematica routine.

The first eight eigenergies coefficients are displayed in Tables 1-2 for
both the considered potentials.

Table1. The coefficients $\epsilon _{n,l}(k)$ in the case of the Cornell
potential for few \ states.

\begin{tabular}{|l|l|l|l|}
\hline
State & $n=0,l=0(1S)$ & $n=1,l=0(2S)$ & $n=0,l=1(1P)$ \\ \hline
$k=0$ & $-\frac{1}{4}$ & $-\frac{1}{16}$ & $-\frac{1}{16}$ \\ \hline
$k=1$ & $3$ & $12$ & $10$ \\ \hline
$k=2$ & $-12$ & $-528$ & $-480$ \\ \hline
$k=3$ & $216$ & $105984$ & $99840$ \\ \hline
$k=4$ & $-6360$ & $-34775040,$ & $-33054720$ \\ \hline
$k=5$ & $245952$ & $14770372608$ & $14051573760$ \\ \hline
$k=6$ & $-11433984$ & $-7410570756096$ & $-7038283284480$ \\ \hline
$k=7$ & $610773696$ & $4197622600433664$ & $3976876146032640$ \\ \hline
\end{tabular}

\bigskip

Table2. The coefficients $4^{k}$ $\epsilon _{n,l}(k)$ in the case of the
quartic anharmonic oscillator for few \ states.

\begin{tabular}{|l|l|l|l|}
\hline
State & $n=0,l=0(1S)$ & $n=1,l=0(2S)$ & $n=0,l=1(1P)$ \\ \hline
$k=0$ & $3$ & $7$ & $5$ \\ \hline
$k=1$ & $15$ & $75$ & $35$ \\ \hline
$k=2$ & $-165$ & $-1575$ & $-525$ \\ \hline
$k=3$ & $3915$ & $66825$ & $16625$ \\ \hline
$k=4$ & $-520485/4$ & $-15184575/4$ & $-2894325/4$ \\ \hline
$k=5$ & $21304485/4$ & $1024977375/4$ & $152440575/4$ \\ \hline
$k=6$ & $-2026946145/8$ & $-155898295875/8$ & $-18353729625/8$ \\ \hline
$k=7$ & $108603230895/8$ & $12977225578125/8$ & $1224596281125/8$ \\ \hline
\end{tabular}

\bigskip

Table3. CPU time (in secunds) for the calculation of $\epsilon _{0,0}(k)$ ($%
k=0,r_{0}$) coefficients in the case of the Cornell potential for the 1S
state.

\begin{tabular}{|l|l|l|l|l|}
\hline
& $r_{0}=10$ & $r_{0}=20$ & $r_{0}=30$ & $r_{0}=40$ \\ \hline
RSPT method & $26.2$ & $188.7$ & $1024.9$ & $2626.2$ \\ \hline
HFHV method & $0.016$ & $0.032$ & $0.062$ & $0.16$ \\ \hline
\end{tabular}

\bigskip

Table4. CPU time (in secunds) for the calculation of $\epsilon _{0,0}(k)$ ($%
k=0,r_{0}$) coefficients in the case of the quartic anharmonic oscillator
for the 1S state.

\begin{tabular}{|l|l|l|l|l|}
\hline
& $r_{0}=10$ & $r_{0}=20$ & $r_{0}=30$ & $r_{0}=40$ \\ \hline
RSPT method & $43.4$ & $277.8$ & $1213.6$ & $2676.6$ \\ \hline
HFHV method & $0.016$ & $0.047$ & $0.11$ & $0.27$ \\ \hline
\end{tabular}

\bigskip 

We have also compared the present method to the usual RSPT one\footnote{%
For the RSPT method, we have implemented, in a Mathematica program, the set
of equations displayed in section 7.3 of \ Ref. \cite{orzagbender}.}, in the
simple case of the 1S state. For both potentials, the respective CPU times%
\footnote{%
Throughout this paper, the mentionned CPU times correspond to runs on a PIV
personnal computer with $1.5$ Go of RAM memory.} are displayed in Tables
3-4. One can see from the latter tables that the computational time of the
HFHV method is \emph{reduced by a factor }$10^{4}$ comparatively to the
usual RSPT method. \ In addition, let us note that the ratio of the
respective CPU time (RSPT/HFHV) grows with increasing order.

Furthermore, the HFHV method reveals itself as a fast and efficient tool for
a systematic calculation of large order perturbative coefficients. Indeed,
considering, for example, the 1S state in the Cornell potential case, the
necessary CPU time to calculate the first 1000 perturbative coefficients is
less than four ($04$) hours\footnote{%
For the other states, the CPU time has the same order of magnitude.}.
Consequently, one can find, in principle, the asymptotic behavior of the $%
\epsilon _{n,l}(k)$ coefficients for large $k$, which reads, for both the
considered potentials:%
\begin{equation}
\epsilon _{n,l}(k)\sim (-1)^{k+1}\text{ }\Gamma \left( k+b(n,l)\right) \text{
\ }a(n,l)^{k}\text{ \ \ for }k\rightarrow \infty \text{ }  \label{asymptotic}
\end{equation}%
where the notation $\Gamma $\ stands for the Euler Gamma function. For
example, for the 1S state, considering the Cornell (resp. quartic
anharmonic) interaction, we have found that $a(0,0)\simeq 6$ (resp. $%
a(0,0)\simeq 3/2$ ) and $b(0,0)\simeq 2$ (resp. $b(0,0)\simeq 3/2$ )%
\footnote{%
These results have been obtained with a satisfactory accuracy from the first
1000 perturbative coefficients. For excited states, one needs more
perturbative coefficients.}. Let us note that, for the quartic anharmonic
potential, the asymptotic behavior (\ref{asymptotic}) together with the
numerical values of $a(0,0)$ and $b(0,0)$\ are in perfect agreement\footnote{%
Up to rescaling of the quantities $g$ and $\epsilon _{n,l}$ to match the
definitions given by Bender and Wu.} with the one dimensional calculation of
Bender and Wu in the framework of WKB analysis and difference-equation
method \cite{benderwu69, benderwu71, benderwu73}. In the Cornell interaction
case, the large $k$ behavior (\ref{asymptotic}) agrees fairly well with the
WKB result \cite{vrscay84}.

Let us make some comments about the asymptotic behavior (\ref{asymptotic})
of the $\epsilon _{n,l}(k)$ coefficients. This large $k$ behavior means
that, for the two potentials under consideration, the expansion (\ref%
{EVformal}) has its radius of convergence equal to zero and is, thus, a\
divergent serie for all $g$. The divergence of the perturbative serie (\ref%
{EVformal}) indicates that taking, for the Cornell (resp. quartic
anharmonic) potential the term $gx$ (resp. $gx^{4}$) as a perturbation leads
to a singular perturbation. For the Cornell (resp. quartic anharmonic)
interaction, a simple way to understand this singular behavior is to
compare, $\exp (-\frac{x}{2(n+l+1)})$ (resp. $\exp (-\frac{x^{2}}{2})$), the
controlling factor of the large $x$ behavior\footnote{%
The notion of controlling factor for large $x$ comes from the fact that $%
x=\infty $ is an irregular singular point of the dimensionessless Schr\"{o}%
dinger\ equation (\ref{sch1}) with potentials (\ref{potU1}) and (\ref{potU2}%
). For more details, see Ref. \cite{orzagbender}.} of the unpertrurbed ($g=0$%
) eigenfunction with, $\exp (-\frac{2g^{1/2}x^{3/2}}{3})$ (resp. $\exp (-%
\frac{g^{1/2}x^{3}}{3})$), the controlling factor of the large $x$ behavior
for $g\neq 0$. One can see an abrupt change in the nature of the
eigenfunction when we go, for large $x$, to the limit $g\rightarrow 0_{+}$.
For the Cornell (resp. quartic anharmonic) potential, this feature occurs
because, for large $x$, the perturbation term $gx$ (resp. $gx^{4}$) is not
negligeable comparatively to $-1/x$ (resp. $x^{2}$).

The other remark deals with the following point. Since the the expansion (%
\ref{EVformal}) is a divergent serie, one may worry about the interest of
the $\epsilon _{n,l}(k)$ coefficients calculation. In fact, different
mathematical methods (like the Pad\'{e} approximants) allow to sum divergent
series and have been used with some success in different fields of physics 
\cite{orzagbender, sumdiv1, sumdiv2, sumdiv3, sumdiv4, sumdiv5, sumdiv6,
sumdiv7, sumdiv8, sumdiv9}. For the eigenergies expansion (\ref{EVformal})
in the case of the considered potentials, the choice of an optimized
summation method is beyond the scope the present work. Anyhow, the interest
of the $\epsilon _{n,l}(k)$ coefficients calculation lies in the fact that
the accuracy of the approximative sum increases with increasing known terms
in the expansion, this feature being independent of the particular choice of
the summation method used.

\section{Comments and conclusion}

Before concluding, some comments are in order. We have shown the uselfulness
of HFHV method for Coulomb plus linear term and the quartic anharmonic
oscillator potentials that interest us for other issues. This method can be
applied, \emph{mutadis mutandis}, to radial interactions of the form $V(r)=$ 
$-1/r+g$ $r^{p}$ ($p\geq 2$) or $V(r)=$ $r^{2}+g$ $r^{2p}$ ($p\geq 3$).

More generally, the HFHV method has been used, in fact, for many
interactions, where the potentials take the form of a power serie (or a
finite polynomial), the unperturbed Hamiltonian being the Coulomb one or the
harmonic oscillator one (for a general discussion see Ref. \cite%
{killingbeck2001}). For instance, the method was applied for: the screened
Coulomb potential \cite{grant79}, the $N-$dimensional generalized
exponential cosine-screened Coulomb potential\footnote{%
This kind of potential is encountered in many fields of physics like plasma,
nuclear and solid state physics.} \cite{chatterjee,sev-tez}, the potential $%
V(r)=$ $r^{2}+\lambda r^{2}/(1+g$ $r^{2})$ \cite{witwit91}, the Gaussian
potential \cite{witwit91b, liolios}, etc. More recently \cite{kwato01}, it
has been applied to a set of potential like the Gaussian potential, Patil
potential, etc. To be complete, let us note that the HFHV method has been
also extended to the relativistic case for the Klein-Gordon equation, with a
Coulomb potential plus polynomial perturbation, to calculate the eigenvalues
perturbative coefficients \cite{vrscay93}.

The interesting and powerful side of the method stems from the fact that one
does not need the eigenfunctions. Indeed, expressing the potential as a
power serie (or a finite polynomial) in the coordinates $x$, the HF and HV
theorems provide recursion relations that give the eigenenergy coefficients
through the knowledge of the expansion coefficients of the moments $\langle
x^{j}\rangle $. The coefficients $x_{j}^{(k)}$ are computed by column, in a
hierarchical manner, beginning with $x_{j}^{(0)}$.

In conclusion, by applying the HF and HV theorems to the interactions (\ref%
{pot1}) and (\ref{pot2}), we have shown that the eigenvalues perturbative
coefficients $\epsilon _{n,l}(k)$ can be obtained, at any order $k$, without
any wave functions calculations. Thus, the HFHV method provides a remarkably
fast and reliable tool to calculate very large order eigenenergies
perturbative coefficients.

\textbf{Acknowledgement }\ We would like to thank S. Hassani for discussions
and critical reading of the manuscript.

\end{document}